\documentstyle[12pt]{article}
\let\ds\displaystyle
\title{Integrability of Riccati equations and the
stationary KdV equations}
\author{R.Z.~Zhdanov \\ \small Institute of Mathematics,\\
\small 3 Tereshchenkivska Street,
252004 Kiev, Ukraine\thanks{e-mail: rzhdanov@apmat.freenet.kiev.ua}}
\date{}
\let\p\partial
\begin{document}
\maketitle
\begin{abstract}
Using the S.Lie's infinitesimal approach we establish the connection between
integrability of a one-parameter family of the Riccati equations and the
stationary KdV hierarchy.
\end{abstract}
In this paper we will suggest a method for integrating a one-parameter family
of the Riccati equations
\begin{equation}
\label{ric}
u_x + u^2 = f(x,\lambda)
\end{equation}
based on their Lie symmetries. Here
\[
f(x,\lambda)=\lambda^n + \lambda^{n-1}V_{n-1}(x)+\cdots+
\lambda V_1(x)+V_0(x)
\]
and $\lambda$ is an arbitrary real parameter.

We recall the principal idea of application of Lie group methods to
integrating the Riccati equation (\ref{ric}). Suppose it admits a
one-parameter transformation group having the infinitesimal operator
\[
X=\xi(x,u,\lambda){\p\over \p x} +
\eta(x,u,\lambda){\p \over \p u}.
\]
Then making a change of variables $(x, u)$ $\to$ $(\tilde x, \tilde u)$
transforming $X$ to become $\tilde X= {\p \over \p \tilde x}$ (which is
always possible) reduces the equation under study to an ordinary differential
equation not containing the independent variable $\tilde x$ (for more details,
see, for example, \cite{olv,ovs}). The latter is evidently integrable by
quadratures which provides integrability of the initial differential equation.
Lie's invariance criterion for the Riccati equation (\ref{ric}) to admit a
one-parameter group having the infinitesimal operator $X$ yields
the following determining equation for $\xi, \eta$:
\[
\eta_x + (\eta_u - \xi_x)(f - u^2) - \xi_u(f - u^2)^2
+ 2u \eta - \xi f_{x}=0.
\]
However integrating the above equation is by no means
easier than integrating the initial equation (\ref{ric}). This
makes an application of the Lie's infinitesimal method in its
full generality inefficient. One should make some additional
guesses about a structure of the operator $X$ enabling to
simplify the determining equation. Our idea is to use a
special Ansatz for the coefficients of the inifinitesimal
operator
\[
\xi=a(x,\lambda),\quad \eta = b(x,\lambda)u + c(x,\lambda),
\]
where $a, b, c$ are polynomials in $\lambda$.

Inserting these expressions into the determining equation
and splitting with respect to $u$ yield
\begin{eqnarray}
&& b + a' = 0,\label{eq1}\\
&&b' + 2c =0,\label{eq2}\\
&&c' + (b - a')f - af' = 0.\label{eq3}
\end{eqnarray}
Hereafter the prime denotes differentiation with respect to $x$.

From (\ref{eq1}), (\ref{eq2}) it follows that
\[
b=-a',\quad c=-\frac{1}{2}a''.
\]

Substituting these expressions into (\ref{eq3}) gives
\begin{equation}
\begin{array}{l}
\label{red}
\frac{1}{2}a''' + 2a'(\lambda^n + \lambda^{n-1}V_{n-1}+\cdots+V_0)\\[2mm]
\phantom{\frac{1}{2}a'''}+a(\lambda^{n-1}V'_{n-1}+\cdots+V'_0)=0.
\end{array}
\end{equation}

Now we fix the following form of an unknown function $a(x,\lambda)$
(see, also \cite{zhd})
\begin{equation}
\label{ansatz}
a(x,\lambda)= \lambda^N + \lambda^{N-1}A_{N-1}(x)+\cdots+
\lambda A_1(x)+A_0(x).
\end{equation}

In what follows we will restrict our considerations to two cases
\begin{enumerate}
\item[{(a)}]{$n$ is arbitrary and $N=1$,}
\item[{(b)}]{$N$ is arbitrary and $n=1$.}
\end{enumerate}

First, we consider the case $N=1$. Inserting $a(x,\lambda)=\lambda + A(x)$
into (\ref{red}) and splitting with respect to the powers of $\lambda$ yield
\begin{eqnarray}
\lambda^n&:& 2A' + V'_{n-1}=0,\nonumber\\
\lambda^j&:& 2A'V_j + AV'_j +V'_{j-1}=0,\quad j=1,\ldots,n-1,\label{rec}\\
\lambda^0&:& \frac{1}{2}A''' + 2A'V_0 + AV'_0=0.\nonumber
\end{eqnarray}

Solving the above recurrent relations (\ref{rec}) we get the following
expressions for $V_j$:
\begin{equation}
\label{gen}
V_{n-j}={j+1\over 2^j}V^j + \sum\limits_{i=1}^{j-1}\,
{i+1\over 2^i}C_{n+i-j}V^i.
\end{equation}
Here $C_0,\ldots,C_{n-2}$ are arbitrary real constants,\
$C_{n-1}\stackrel{\rm def}{=}0$ and $V=V(x)$ is a solution
of the third-order ODE:
\begin{equation}
\label{gkdv}
\frac{1}{4}V''' + {(n+1)(n+2)\over 2^{n+1}}V^nV'
+ \sum\limits_{i=1}^{j-1}\, {(i+1)(i+2)\over 2^{i+1}}C_{i}V^iV'+ C_0V'=0.
\end{equation}
The above equation is evidently integrated by quadratures which
means that given the conditions (\ref{gen}), (\ref{gkdv}) the
Riccati equation (\ref{ric}) is integrable by quadratures. Note
that the equation (\ref{gkdv}) with $C_1=\cdots=C_{n-2}=0$ and $n=1$
is nothing else than the standard stationary KdV equation
\begin{equation}
\label{kdv}
\frac{1}{4}V''' + {3\over 2}VV' + C_0V'=0.
\end{equation}
Choosing $C_1=\cdots=C_{n-2}=0$ and $n=2$ yields the stationary
modified KdV equation
\begin{equation}
\label{mkdv}
\frac{1}{4}V''' + {3\over 2}V^2V' + C_0V'=0.
\end{equation}

Now we will turn to the case of an arbitrary $N$ with $n=1$.
With this choice of $N, n$ equation (\ref{red}) takes the form
\[
\frac{1}{2}\sum_{j=0}^N\, A'''_j\lambda^j + 2\sum_{j=0}^N\,A'_j\lambda^j
(\lambda + V_0) + \sum_{j=0}^N\, A_j\lambda^j V'_0=0
\]
with $A_n=1$.
Splitting the above equality with respect to the powers of $\lambda$
yields recurrent relations for $A_j$
\begin{equation}
\label{rkdv}
\frac{1}{2} A'''_j + 2V_0 A'_j +  A_j V'_0 + 2A'_{j-1}=0,\quad j=0,\ldots,N
\end{equation}
with $A_{-1}\stackrel{\rm def}{=}0$.

The first $N$ relations of (\ref{rkdv}) are solved by subsequent
integrations yielding the expressions for the functions
$A_0(x),\ldots,A_{N-1}(x)$ via the function $V_0(x)$ and its
derivatives. Substituting these results into the last equation ($j=0$)
we arrive at nonlinear ordinary differential equation
for the function $V_0(x)$. To reveal the structure of the equation in
question we introduce the new functions ${\cal U}_0(x)$,\ ${\cal
  U}_1(x)$,$\ldots$ by the following recurrence relation:
\begin{equation}
\label{den}
  {\cal U}_{j}(x)=\left(\underbrace{-{\ds\frac{1}{4}}{\ds{d^2\over dx^2}} - V_0(x) +
    {\ds\frac{1}{2}}\left ({\ds{d\over dx}}\right )^{-1}V_0'(x)}_{{\cal R}}\right)\,
  {\cal U}_{j-1},\quad j=0,1,\ldots,
\end{equation}
where ${\cal U}_{-1}(x) \stackrel{\rm def}{=}1$.

Being so determined the functions ${\cal U}_j(x)$ are easily recognized
to be the conserved densities for the KdV equation. Furthermore, the
integro-dif\-fe\-ren\-ti\-al operator ${\cal R}$ is the recursion operator whose
repeated action on some initial conserved density yields the whole
hierarchy of the conserved densities for the KdV equation (for more details
on recursive operators for the KdV equation, see e.g. \cite{olv,ibr,abl}).

Now we can solve the first $N$ relations of (\ref{rkdv}) in terms of
the functions ${\cal U}_j(x)$
\begin{equation}
\label{hie}
A_{N-j}(x)={\cal U}_{j-1}+\sum\limits_{k=1}^{j-1}\,C_{N-k}\, {\cal U}_{j-k-1}(x) +
C_{N-j},\quad j=1,\ldots,N,
\end{equation}
where $C_0,\ldots,C_{N-1}$ are integration constants.

As $A_{-1}\stackrel{\rm def}{=}0$, the last equation from (\ref{rkdv}) can be
rewritten to become
\[
D_x {\cal R} A_0=0
\]
or, equivalently,
\begin{equation}
\label{last}
D_x\left(\sum\limits_{j=0}^{N-1}\,C_{j}\,
{\cal R}^{j} + {\cal R}^{N}\right){\cal U}_0=0,
\end{equation}
where ${\cal U}_0=-\frac{1}{2}V$. In what follows we will show that this
equation is nothing else than the stationary higher KdV equation.
To this end we will need the following operator identity:
\[
D_x\,({\cal R}^j)\equiv \left (D_x {\cal R}D_x^{-1}\right )^j\,D_x.
\]
The integro-differential operator
\[
{\cal P}=D_x {\cal R}D_x^{-1}=-\frac{1}{4}D_x^2 - V - \frac{1}{2}V'D_x^{-1}
\]
is the second recursion operator for the KdV equation. Acting repeatedly
with ${\cal P}$ on some initial symmetry (say ${\cal F}_0=-\frac{1}{2}V_x$)
yields the whole hierarchy of higher symmetries of the KdV equation\
${\cal F}_1, {\cal F}_2,\ldots$, where
\[
{\cal F}_j={\cal P}^j\,{\cal F}_0,\quad j\ge 1.
\]

In view of the above facts we can represent equation (\ref{last}) in the
following form:
\begin{equation}
\label{hkdv1}
\left(\sum\limits_{j=0}^{N-1}\, C_j {\cal P}^{j} +
{\cal P}^{N}\right)\,D_x{\cal U}_0 = 0.
\end{equation}

Taking into account the fact that $D_x{\cal U}_0 = -\frac{1}{2}V' = {\cal F}_0$
we get finally
\begin{equation}
\label{hkdv2}
\sum\limits_{j=0}^{N-1}\, C_j {\cal F}_{j} +
{\cal F}_{N} = 0
\end{equation}
which is the standard form of the stationary higher KdV equation.
Provided $N=1$, it coincides with the stationary KdV equation (\ref{kdv}).

As the stationary higher KdV equations are integrable \cite{nov}, the
initial Riccati equation (\ref{ric}) with $N=1$ is integrable by
quadratures provided $V(x)$ is a solution of one of the equations of
the stationary KdV hierarchy.

\end{document}